\documentclass[aps,pre,showpacs,floats,twocolumn,amsfonts,amssymb,unsortedaddress,floatfix]{revtex4}
\usepackage{bm}
\usepackage{amsmath}
\usepackage{graphicx}

\begin{document}

\addtolength{\topmargin}{10pt}

\title{Annealing a Magnetic Cactus into Phyllotaxis}

\author{Cristiano Nisoli$^{1}$, Nathaniel M. Gabor$^{2}$, Paul E. Lammert$^{3}$, J. D. Maynard$^{3}$ and Vincent H. Crespi$^{3}$.}

\affiliation{$^{1}$ Theoretical Division and Center for Nonlinear Studies,  Los Alamos National Laboratory, Los Alamos NM 87545 \\ 
$^{2}$Department of Physics 
Cornell University, 109 Clark Hall, Ithaca, NY 14853-2501 \\ 
$^{3}$Department of Physics 
The Pennsylvania State University, University Park, PA 16802-6300}

\date{\today}

\begin{abstract}
The appearance of  mathematical regularities in the disposition of leaves on a stem, scales on a pine-cone and spines on a cactus has puzzled scholars for millennia; similar so-called phyllotactic patterns are seen in self-organized growth, polypeptides, convection, magnetic flux lattices and ion beams. Levitov showed that a cylindrical lattice of repulsive particles can reproduce phyllotaxis under the (unproved) assumption that minimum of energy would be achieved by 2-D Bravais lattices. Here we provide experimental and numerical evidence that the Phyllotactic lattice is actually a ground state.  When mechanically annealed, our experimental ``magnetic cactus''  precisely reproduces botanical phyllotaxis, along with domain boundaries (called transitions in Botany) between different phyllotactic patterns.  We employ a structural genetic algorithm to explore the more general axially unconstrained case, which reveals multijugate (multiple spirals) as well as monojugate (single spiral) phyllotaxis. 
\end{abstract}
\maketitle 

\section{Introduction}

\begin{figure}[t!]
\begin{center}
\includegraphics[width=3 in]{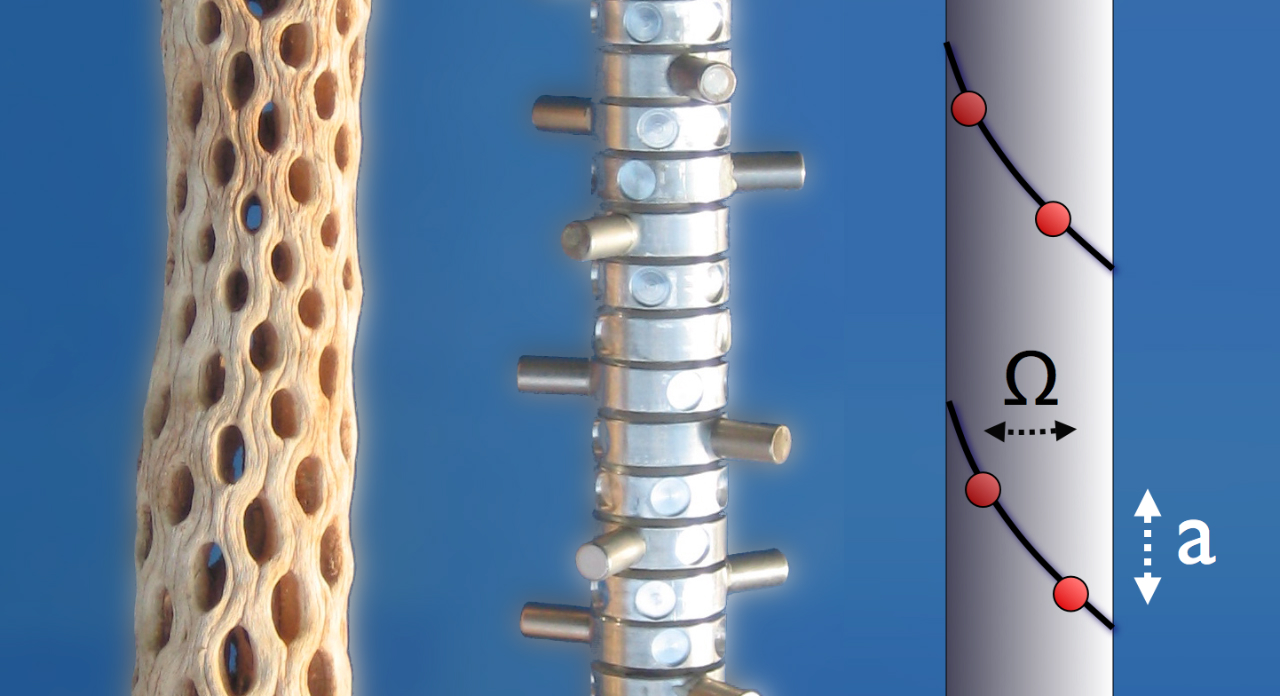} 
\caption{Natural and Magnetic Cacti. A specimen of {\it Mammillaria elongata} displaying a helical morphology ubiquitous to nature, a magnetic cactus of dipoles on stacked bearings, and a schematic of a wrapped Bravais lattice showing the angular offset (divergence  angle) $\Omega$ and the axial separation $a$ between particles.}
\label{Fig1}
\end{center}
\end{figure}

Symmetrical morphologies and regular patterns in living organisms (Fig~\ref{Fig1}) have been credited with originating the idea of beauty, the notion of art as an imitation of nature, and humanity's first mathematical inquiries~\cite{Adler,Jean,Smith,Grew}. The fascinating symmetrical patterns of organs in plants, called phyllotaxis~\cite{Adler,Jean, Smith}, were known to the Romans (Pliny) and ancient Greeks (Theofrastus), while early recognitions are found in sources as ancient as the Text of the Pyramids~\cite{Adler}. Leonardo da Vinci~\cite{daVinci}, Andrea Cesalpino, and Charles Bonnet~\cite{Bonnet} studied phyllotaxis in the modern era. Kepler proposed that the Fibonacci sequence (1, 2, 3, 5, 8\dots), where each term is the sum of the two preceeding ones~\cite{Fibonacci}, describes these phyllotactic patterns. 

\begin{figure}[t!]
\begin{center}
\includegraphics[width=2.8 in]{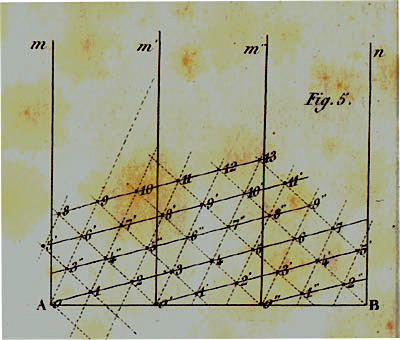} 
\caption{The Bravais lattice with cylindrical boundary conditions that defines a phyllotactic spiral. The cylinder axis is vertical, while the horizontal direction contains three circumferential repeats. The solid line is the generative spiral: this one-dimensional Bravais lattice generates the full structure. The dashed lines are the so-called parastichies or visible secondary spirals: they connect nearest neighbors on the surface of the cylinder. Adapted from A. Bravais and L. Bravais, 1837~\cite{Bravais}.}
\label{Lattice}
\end{center}
\end{figure}

A discipline that  thrived on multidisciplinary interactions~\cite{multi}, phyllotaxis found its standard mathematical description when August and Louis Bravais~\cite{Bravais} introduced the point lattice on a cylinder to represent the dispositions of leaves in 1837 (see Fig.~\ref{Lattice}), thirteen years {\it before} August's seminal work on crystallography~\cite{Bravais2}. Unfortunately botanists neglected the work of the Bravais  brothers, and it wasn't until Church  rediscovered it eighty years later that more progress was achieved  in the field~\cite{Church}. 




The geometrical description of cylindrical phyllotaxis relies, in the simplest case, on the phyllotactic lattice introduced by the Bravais brothers~\cite{Adler,Jean,Smith,Bravais}. It consists of a so-called generative spiral of divergence angle $\Omega$. We can visually decompose the resulting lattice in crossing spirals  that join nearest neighbors, as in Fig.~\ref{Lattice},  which botanists call 
parastichies. It is a fundamental observation (made first by Kepler) that the numbers $n$, $m$ of crossing parastichies  needed to cover the lattice are consecutive terms of the standard Fibonacci sequence, or less frequently the variants obtained by changing the second term, also called Lucas numbers: 1, 3, 4, 7, 11\dots  and 1, 4, 5, 9\dots often referred to as  second and third phyllotaxis. From that, one can prove that the divergence angle of the generative spirals in plants assumes values close to~\cite{Jean,Adler74}
\begin{equation}
\Omega_p=\frac{360 ^\circ}{\left(\tau +p\right)}, 
\label{Omega}
\end{equation} 
where $p=1, 2, 3$ denotes first, second or third phyllotaxis and $\tau=\left(1+\sqrt{5}\right)/2$ is the golden ratio. For more than one generative spiral (``multijugate'' phyllotaxis), parastichies share a common divisor  $(n,m)=(k n',k m')$, $k$ being the number of generative spirals~\cite{Adler74,Jean}. Not unlike domain boundaries in crystals, plants show kinks between domains, called transitions by botanists~\cite{kink,Adler}. 


In the last 50 years, phyllotactic  patterns have been seen or predicted outside of botany: polypeptide chains~\cite{Frey, Abdulnur}, tubular packings of spheres~\cite{Erickson}, convection cells~\cite{Rivier}, layered superconductors~\cite{Levitov}, self-assembled microstructures~\cite{Chaorong}, and cooled particle beams~\cite{Rahman,Shatz}. While it is still debated whether such systems might shed light on botanical phyllotaxis, the occurrence of such mathematical regularities outside of botany is fascinating and leads to generalizations that -- unlike quasistatic botany -- allows for dynamics~\cite{Nisoli_PRL}. 

In a groundbreaking work Levitov recognized phyllotaxis in vortices of layered superconductors~\cite{Levitov}. He next described how phyllotactic patterns represent states of minimal energy of a cylindrical lattice (that is of a lattice with cylindrical boundary conditions) of mutually repelling objects, the repulsion mimicking the interactions between spines, leaves, or seeds in plant morphology~\cite{Levitov2, Levitov3}.  Yet such a constraint to a lattice is absent both in botany and in the physical systems to which this energetic model might apply,  such as adatoms or low-density electrons on nanotubes and ions or dipolar molecules in cylindrical traps. 

Following up on earlier work that focused on the dynamics of rotons and solitons in physical phyllotactic systems~\cite{Nisoli_PRL}, we provide here a detailed experimental and numerical demonstration that Levitov's constraint is not necessary, and that the lowest energy states of repulsive particles in cylindrical geometries are indeed phyllotactic lattices.  In addition, we describe the experimental and numerical generation of multijugate phyllotaxis, static kink-like domain boundaries between different phyllotactic lattices, and unusual disordered yet reflection-symmetric structures that may be a static relic of soliton propagation. 

We show that when a ``magnetic cactus'' of magnets (spines) equally spaced on co-axial bearings (stem) with south poles all pointing outward is annealed, it precisely reproduces botanical phyllotaxis. When studied numerically via a structural genetic algorithm, the fully unconstrained case reveals both multijugate and mono   jugate phyllotaxis. In addition to our macro-scale implementation, such systems could also be created at the quantum level in nanotubes or cold atomic gases.

In section II we describe the statics of repulsive particles in cylindrical geometries. In section III we detail the experiment on the magnetic cactus. In Section IV we discuss the more general case of multijugate phyllotaxis.

\section{Phyllotaxis of repulsive particles in cylindrical geometries}

In this section we will recall Levitov's model~\cite{Levitov2, Levitov3} and some of our own findings~\cite{Nisoli_PRL}. Following Levitov, let us assume that the lowest energy configuration for a set of particles with repulsive interactions,  confined to a cylindrical shell of radius $R$, is a helix with a fixed angular offset $\Omega$ between consecutive particles and a uniform axial spacing $a$, as in Fig.~\ref{Fig1} (this so far unproved ansatz will be investigated later both numerically and experimentally). For a generic pair-wise repulsive interaction $v_{ij}$ between particles $i$ and $j$, the energy of the helix is $V = \frac{1}{2}\sum_{i \neq j} v_{i,j}$. Since the lattice structure is defined by $\Omega$, we can write $V(\Omega)$. 

\begin{figure}[t!!!!!!!!!!!!!!!]
\center
\includegraphics[width=3.2 in]{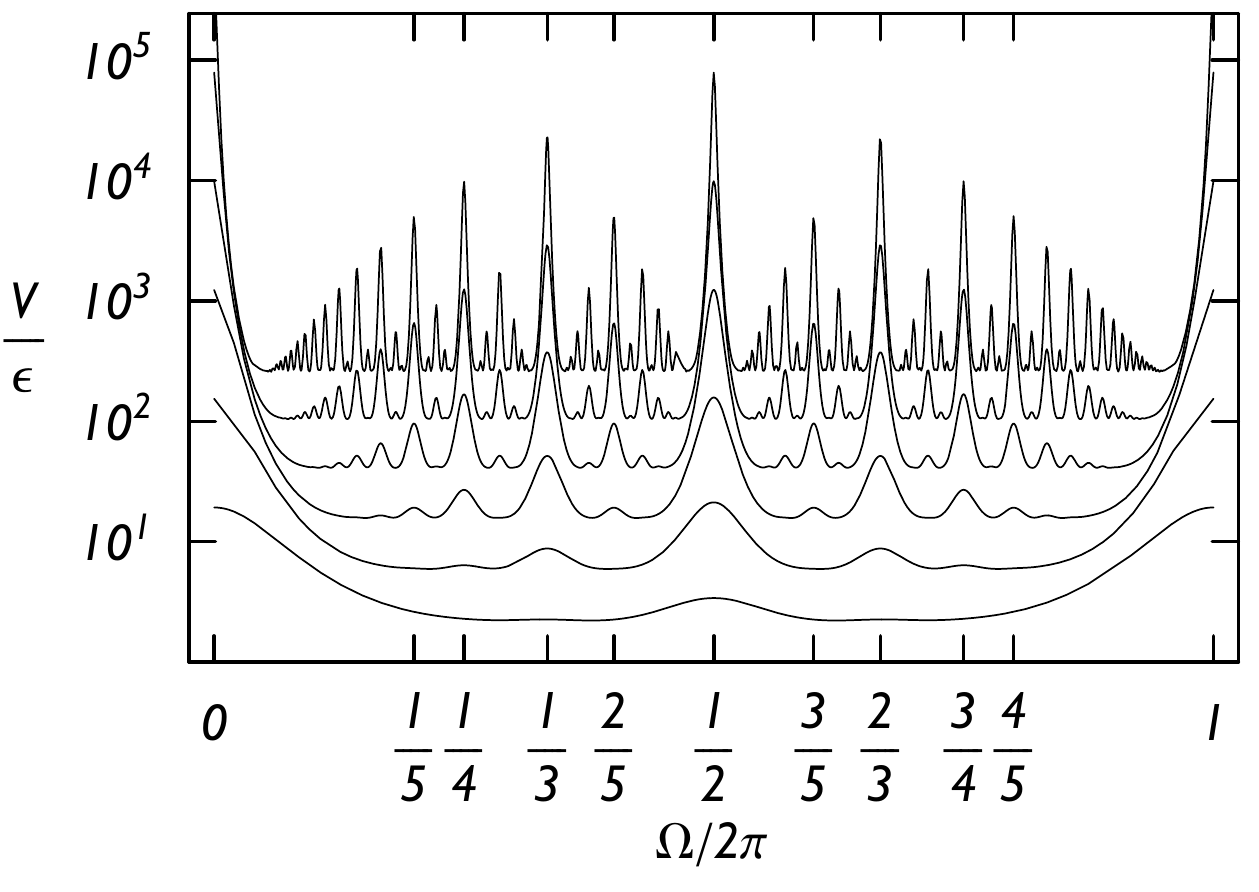}
\vspace{-2mm}
\caption{Lattice energy $V(\Omega)$ versus divergence angle for successively halving values of $a/R$ starting from $0.5$  (using dipole dipole interaction, $\epsilon=p^2/a^2$, where $p$ is the magnetic dipole). Notice the proliferation of peaks as $a/R$ decreases. Reproduced from \cite{Nisoli_PRL}.}
\label{spectra}
\vspace{-2mm}
\end{figure}

In Fig.~\ref{spectra} we plot $V\left(\Omega\right)$ for various values of the ratio $a/R$: 
for specificity we employed  a dipole dipole interaction
 $v_{i,j}={\bm p}_i \cdot {\bm p}_j/ r_{i,j}^3 -3 ({\bm p}_i \cdot {\bm r_{i,j}})({\bm p}_j\cdot {\bm r_{i,j}}) /r_{i,j} ^5 $, repulsive at the densities considered here. However, the following considerations only depend upon geometry and therefore apply to a vast range of 
reasonably behaved, long range repulsive interactions. 

When  $a/R \gg 1$,  the angle $\Omega = \pi$ maximizes distance between neighboring particles and therefore $V\left(\Omega \right)$ has a minimum in $\pi$. The angle between second nearest neighbors along the helix is $2\pi$, which means that they face each other. 
And thus, as the density increases, interaction between the facing second nearest neighbors becomes predominant, and $\Omega = \pi$ is not a minimum for $V\left(\Omega \right)$ anymore. If whe shift the helical angle from $\pi$, the repulsive interaction between second nearest neighbors is reduced, with minimal penalty from nearest neighbors. In terms of $V\left(\Omega \right)$, that means a local maximum $\Omega=\pi$. 

This argument can be iterated for every commensurate winding that allows particles separated by $j$ neighbors to face each other. As density increases further, the angles $2\pi/3$ and $4\pi/3$ also become unfavorable due to third-neighbor interactions. Any commensurate spiral of divergence angle $\Omega=2\pi i/j$ with $i, j$ relatively prime corresponds to a configuration where each particle faces each $jth$ neighbor. For every $j$ there will be a value of $a/R$ low enough such that $\Omega=2\pi i/j$ is a local maximum, which we call a peak of rank $j$. 

The proliferation of peaks for increasing linear density is shown in Fig~\ref{spectra}. We can see that at high density, peaks of equal rank are nearly degenerate; that is natural, since their principal defining energetic contribution arises from particles facing each other at a distance $j a$. The minima also become more nearly degenerate as the density increases. That can be explained intuitively, since for angles incommensurate to $\pi$ each particle ``sees'' the others as incommensurately smeared around the cylinder, and is therefore embedded in a nearly uniform background charge from the other particles. The degenerate energy of the ground state can be well approximated by an uniform continuum distribution $\epsilon_0$, whereas the energy of a peak of order $j$ will be $V\left(2\pi i/j\right)\simeq v(j a)+\epsilon_0$, where $v(r)$ is the energy of two particles facing at a distance $r$: for our dipole interaction $v(ja)=p^2/a^3 j^3$.

The first step to calculate the degeneracy of our system at a given density, is to find the corresponding maximum rank of the peaks. As all of the peaks of the same rank have the same energy, and appear in the spectrum together, we can focus on the emergence of $2\pi /J$. For $a/R\ll1$, this  new peak will emerge when the distance between particles separated by a distance $Ja$  equals that of particles separated by $2\pi R/J$. Therefore one finds for the maximum rank
\begin{equation}
J = \Big\lbrack{\kern  -0.1 em}\Big\lbrack 
\sqrt{\frac{2\pi R}{a}} \Big\rbrack{\kern -0.12 em}\Big\rbrack,
\label{J}
\end{equation}
which as expected only depends on purely geometrical parameters. A little 
more tricky is to compute the degeneracy, given $J$. The set of all the peaks has the cardinality of the class of all the fractions $i/j$, with $i,~j$  coprime and $j\le J$.  This can be considered as the disjointed union of other classes, called Farey classes of order j, defined as follows:  $P_j\equiv \{\Omega = 2\pi i/j\mid$  for $i, \!\ j$ coprime and $i\leq j\}$, i.e. all fractions in lowest terms between 0 and 1 whose denominators do not exceed $j$~\cite{Farey}. The union of all Farey classes up to a certain order $J$ has the cardinality of the set of peaks for a spectrum of maximum rank $J$. Now, the cardinality of $P_j$ is know from number theory to be Euler's totient 
function, $\phi (j)$~\cite{Totient}. Therefore, the degeneracy $D$ of the energy minima for a system with a maximum peak rank $J$ is~\cite{Totient}
\begin{equation}
D=\sum_{j=1}^{J}\phi(j)=\frac{3}{\pi^2} J^2 +O\left(J \log J\right),
\end{equation}
which, from Eq.~\ref{J},  scales as $D\sim 2R/a$.

Finally, we recall~\cite{Smith,Levitov} that the order $j_1$, $j_2$ of the peaks bracketing a minimum relates to its structure in a straighforward way:  the helix corresponds to a rhombic lattice where each particle has its nearest neighbors at axial displacements of $\pm a j_1$, $\pm a j_2$ and second nearest neighbors at $\pm a (j_1+j_2)$ or $\pm a (j_1-j_2)$~\cite{Levitov}.  Also, $j_1$ and $j_2$ give the number of crossing secondary spirals (parastichies) needed to cover the lattice by connecting nearest neighbors. 

For completeness, let us now follow Levitov~\cite{Levitov,Levitov2,Levitov3}, and consider the adiabatic evolution of our system as the linear density is increased. As new sets of maxima and minima emerge,
 the true minimum goes through a series of quasi-bifurcations, the consequence of an elusive symmetry whose explanation goes beyond our scope. 
Suffice it to say that the system evolves quasi-statically from one of these optimal $\Omega$ to another as $R/a$ increases, asymptoting to the golden angle  $\Omega_1=2\pi/\left(\tau +1\right)$ $\left[\tau=\left(1+\sqrt{5}\right)/2\right]$, ubiquitous in botany, as each minimum is bracketed by peaks whose ranks, because of the Farey tree structure described above, are consecutive elements of the Fibonacci sequence.  Occasional ``wrong turns'' at later stages, will not shift the convergence too far from the golden angle, yet the Fibonacci structure would be lost. However if one or two consecutive wrong turns happen at the second or second and third bifurcations the system will converge to the alternative angles of second or third phyllotaxis, given by Eq.~(\ref{Omega}).

We have only surveyed so far spiraling lattices generated by a single helix. A straightforward generalization gives multijugate phyllotaxis, when two or more elements grow at the same axial coordinate~\cite{Adler, Jean, Smith}. This case, which Levitov does not explore, can be easily mathematically reduced to monojugate case, by considering two or more replicas of the phyllotactic lattice as in Fig.~\ref{Lattice}. In our experimental realization we restrict ourselves to the monojugate phyllotaxis, and explore multijugate only numerically.

\section{A Magnetic Cactus}

There is a long history of experimental reproductions of phyllotactic patterns. Recently, Doady {\it et al.} described phyllotaxis in terms of dynamical systems and then verified it experimentally by examining dynamical processes in droplets of ferro-fluid~\cite{Douady}. But even  more than a century ago, Airy showed that phyllotaxis emerged in optimal packing of hard spheres connected by a rubber band, once the band was twisted to increase density~\cite{Airy}. 

Here we expand on what was announced in a recently published Letter~\cite{Nisoli_PRL}: we verify experimentally the assumptions of Levitov's energetic model, by studying  the low energy configurations of interacting magnets stacked evenly-spaced and free to rotate around a common axis. We constructed a mechanical system that  it is free to explore the three angles of botanical phyllotaxis (Eq.~\ref{Omega}).

\subsection{Experimental Apparatus}

\begin{figure}[t!!!!!!]
\center
\vspace{3 mm}\includegraphics[width=2.7 in]{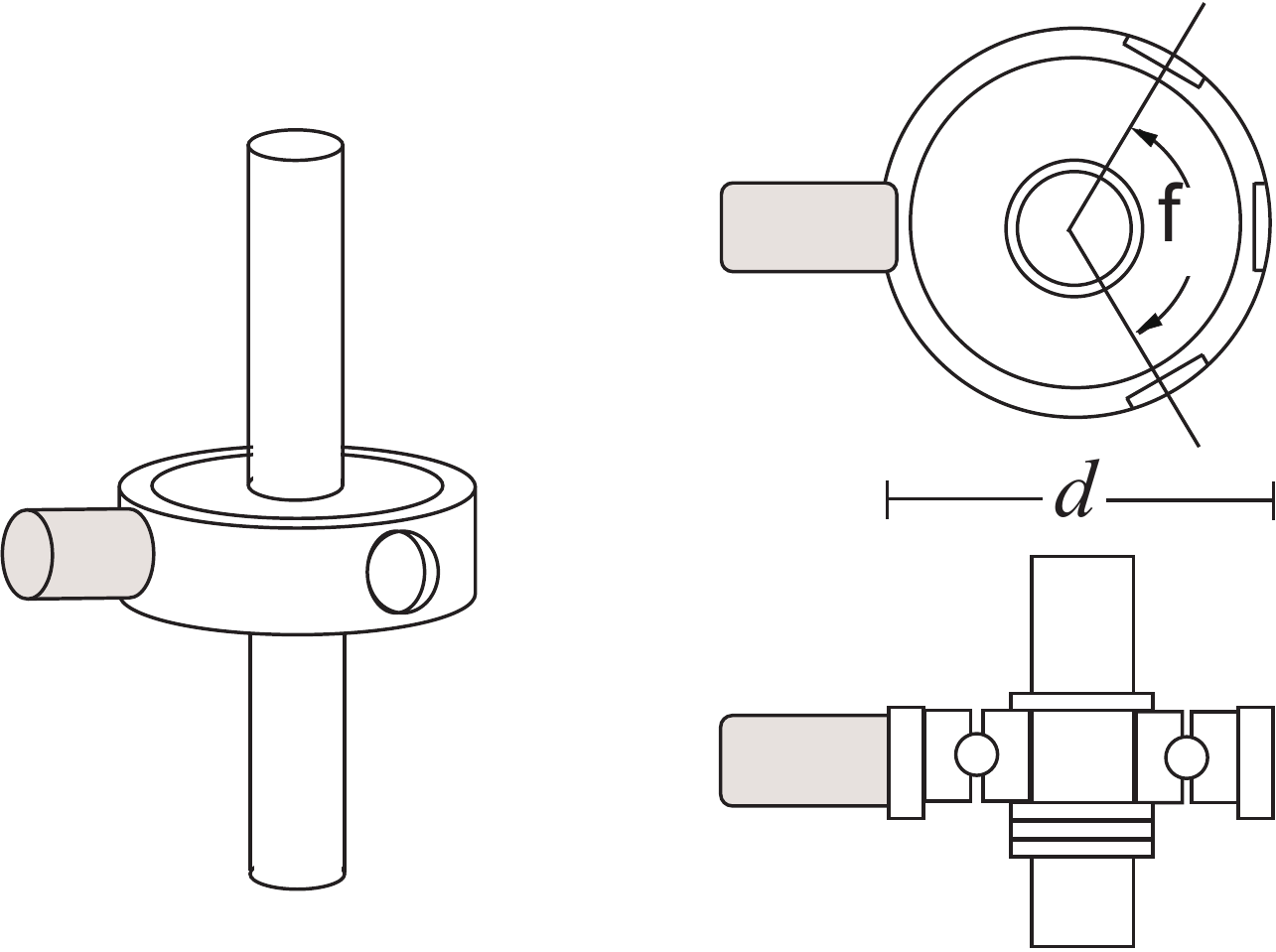}\vspace{10 mm}

\includegraphics[width=2.9 in]{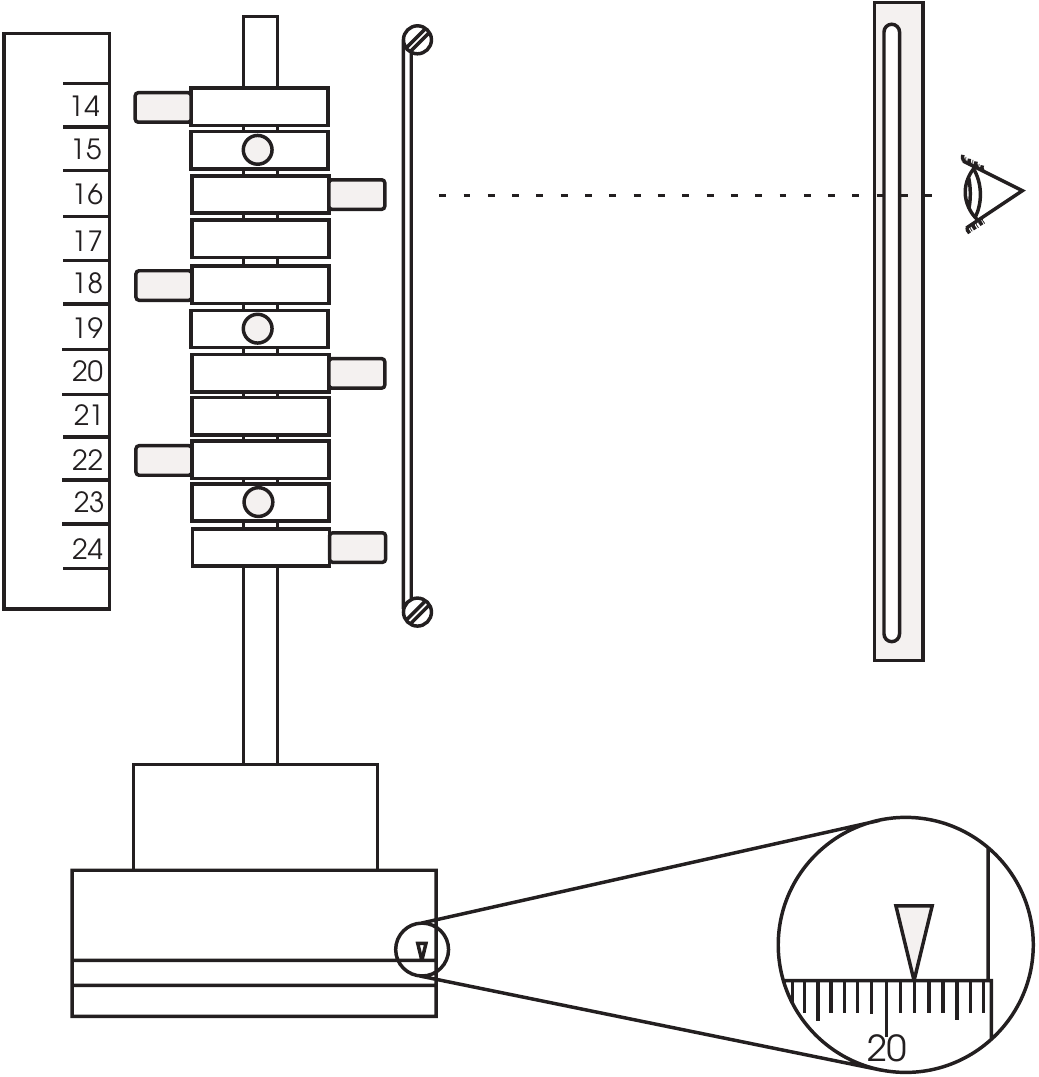}
\caption{Experimental apparatus. Top: Each unit of the magnetic cactus consists of a magnet element and a unit ring secured to a central axis.  The ring diameter \emph{d} is 2.2 cm.  Bottom: a schematic representation of the mounted
magnetic cactus and surrounding measurement devices.  The viewer's eye is restricted by the viewing slit and the reference wires.  Measurements are taken
directly from the dividing head.}\vspace{-12mm}
\label{apparatus}
\end{figure}

We built a  magnetic cactus by mounting permanent magnets (spines) on stacked co-axial bearings (a stem) which are free to rotate about a central axis, as in Fig.~\ref{apparatus}. All the magnets point outward, to produce a repulsive interaction between all magnet pairs. To avoid effects of gravity, the apparatus rests in the vertical position, and is non-magnetic.  We built two different versions, the second with magnets twice as long as the first, as to have a larger effective radius  which gives  three rather than two stable structures. 

We employed cylindrical permanent iron-neodymium magnets, 1.2 cm long and 0.6 cm in diameter. They are mounted on fifty aluminum rings of 2.2 cm outer diameter, each affixed to a non-magnetic radial ball bearing (acetal/silicon, Nordex) as in Fig.~\ref{apparatus}. These unit rings are evenly spaced on an aluminum rod in a stacked structure of 39.9 cm axial length. 

At static equilibrium  we measure separation angle between each magnet element, by rotating the cactus until a magnet element aligns with the reference wires. 
 A telescope and a vertical viewing slit accompanied by two vertical reference wires assist in data acquisition. 
 
\begin{figure}[t!]
\center
\hspace{10 mm}\includegraphics[width=2.2 in]{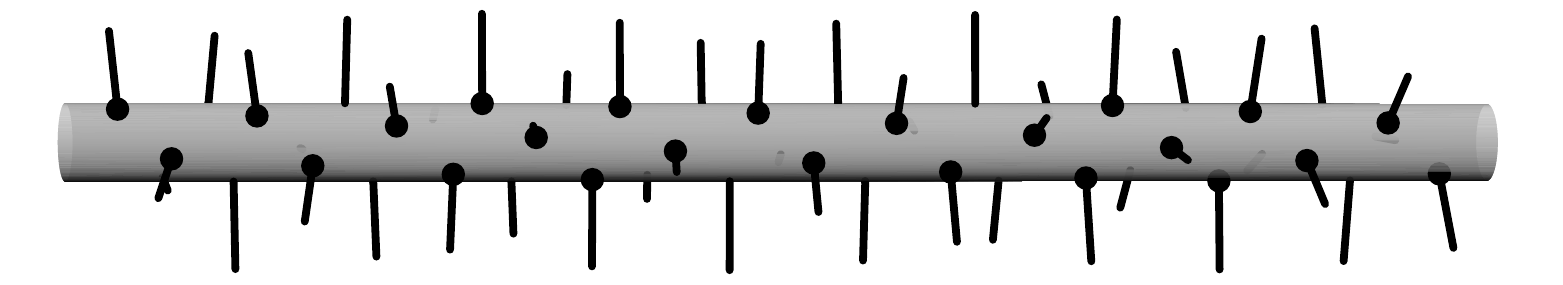}

\hspace{10 mm}\includegraphics[width=2.2 in]{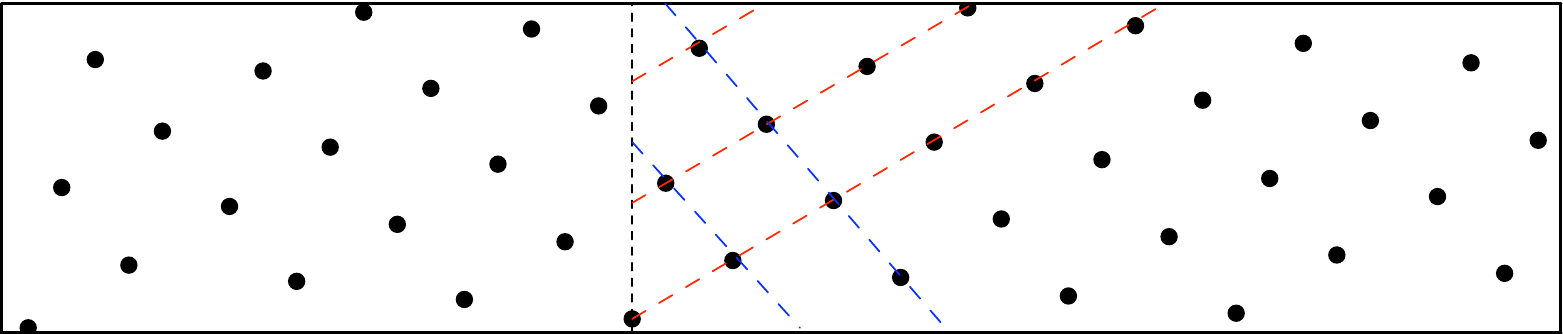}\vspace{3 mm}

\includegraphics[width=2.9 in]{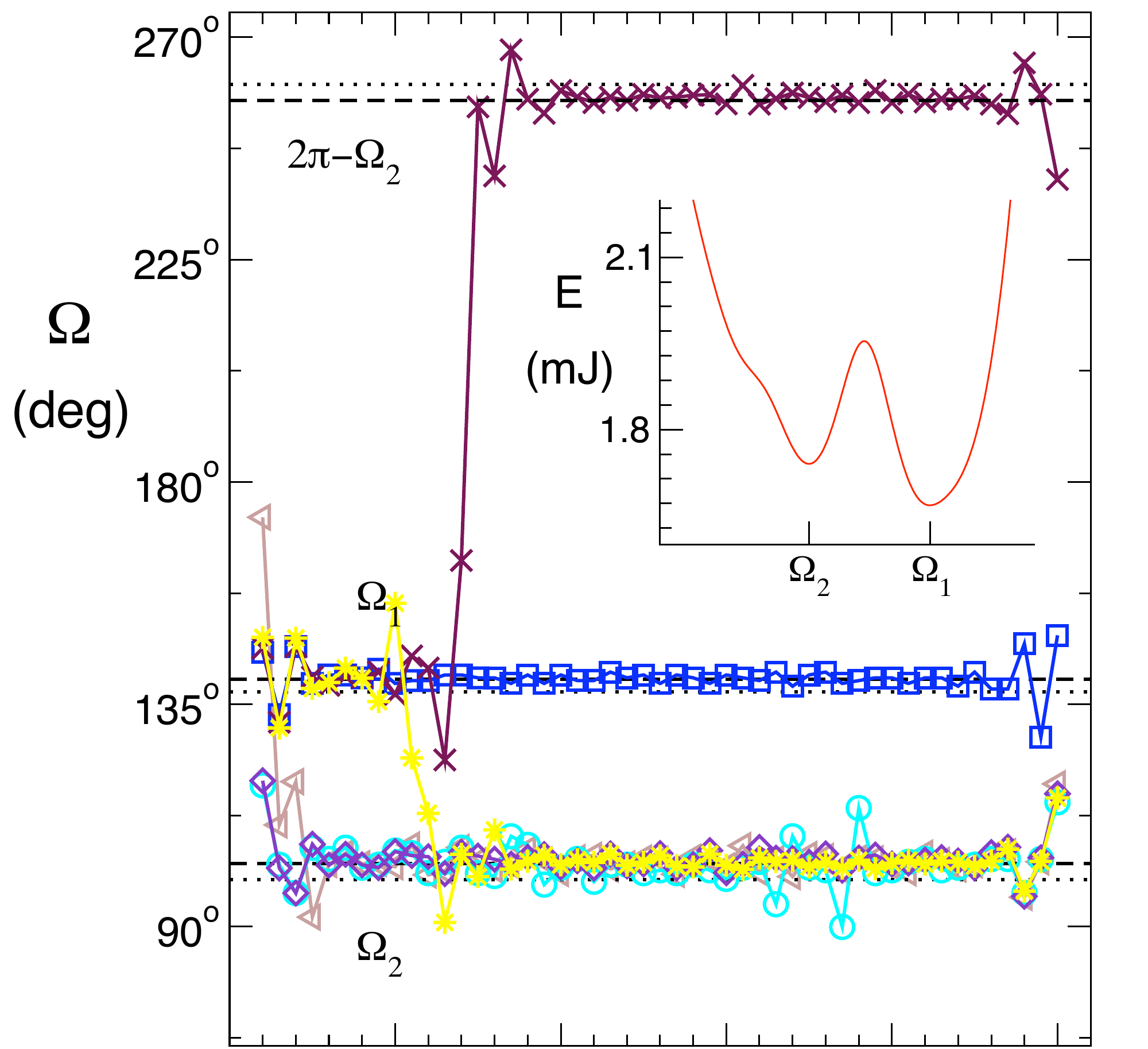}\vspace{-.5 mm}
\includegraphics[width=2.9 in]{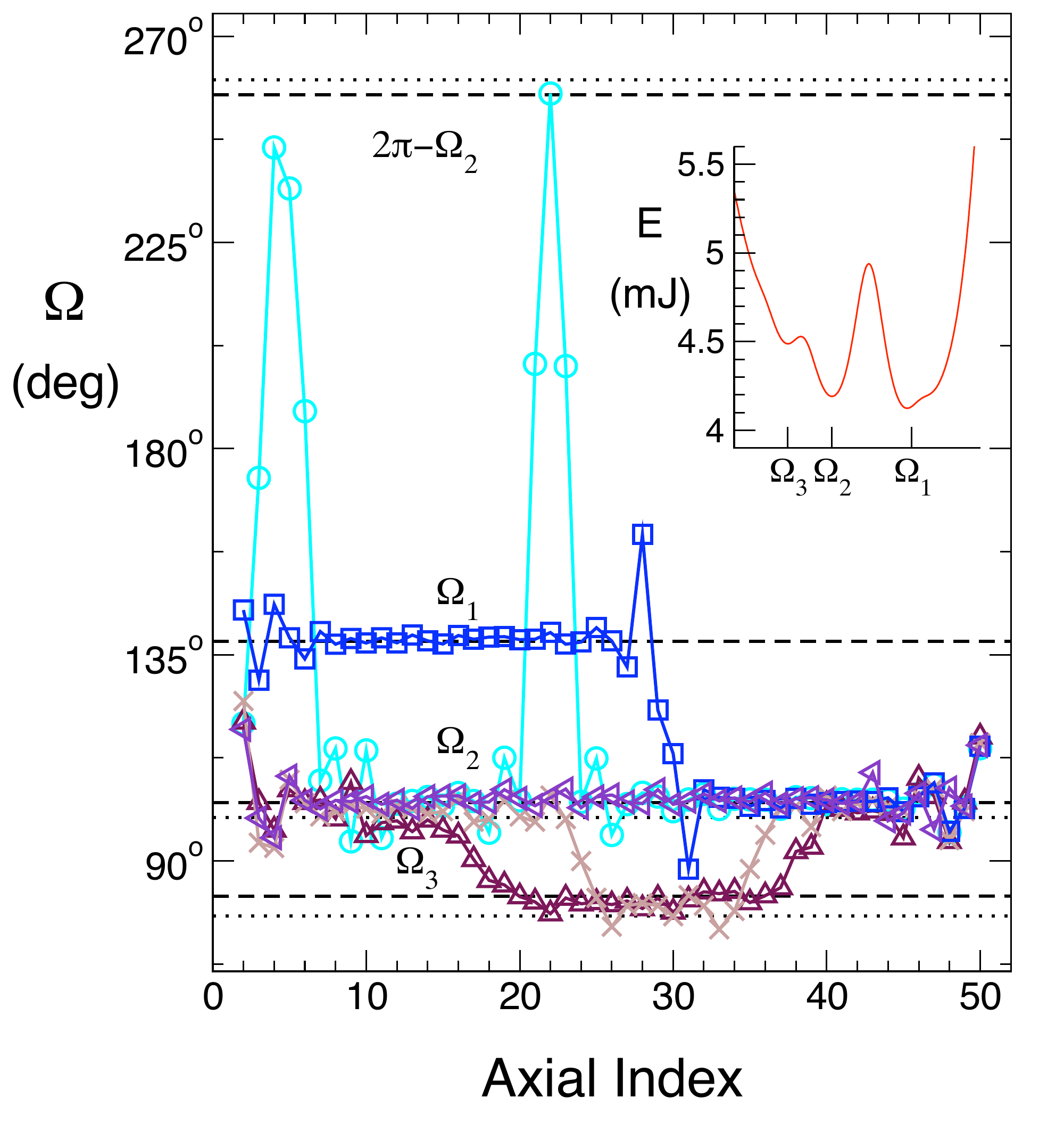}
\caption{Top: A 3-D  rendering of the experimental data and  the corresponding Bravais lattice of the magnetic cactus annealed in a spiral configuration of divergence angle $\Omega_1$ and parastichies (2,3) (blue and red dashed lines), Fibonacci numbers. The bottom two panels show the experimentally measured angular offsets $\Omega$ between successive magnets for magnetic cacti with short (middle) and long (bottom) magnets, plotted versus magnet index, which simply counts the number of magnets along the axis. Flat regions are perfect spirals while steps are boundaries between different phyllotactic domains. The dotted lines give the phyllotactic angles $\Omega_1$, $\Omega_2$, $\Omega_3$ and $2\pi-\Omega_2$ defined in the text, whereas the dashed lines are minima of the magnetic lattice energy (insets) calculated by interpolating the measured pair-wise magnet-magnet interaction.  Data reproduced from Ref. \cite{Nisoli_PRL}.}
\label{Fig2}
\end{figure}

\subsection{Annealing}

By measuring the dipole-dipole interaction between an individual magnet pair, we can reconstruct the curve of the lattice energy $V(\Omega)$ as a function of the angular offset $\Omega$ between magnets. We find that the first arrangement, with short magnets, admits two minima, given by the angles of Eq.~\ref{Omega} for $p=1,2$. The second arrangement, with long magnets, has three minima corresponding to the angles of Eq.~\ref{Omega} $p=1,2,3$, one of which ($p=3$) is a weak metastable minimum. These divergence angles of stable helices  are very close to those predicted by phyllotaxis, of Eq.~\ref{Omega}, and are all accessible by  experimental procedure described below. 

Before every data acquisition, the  cactus is disordered and then athermally annealed into a low-energy state.  The protocol involves repeatedly winding the bottom-most magnet to generate an  ever-tightening spiral, until an explosive release of energy disorders the lattice. Next, an independent external magnet is oscillated in small circular motions near randomly chosen points while the cylinder as a whole is slowly rotated, to further randomize magnet orientations. After  10-30 second of mechanical annealing through applied vibrations, the system consistently enters a robust ordered state which does not anneal further on experimental timescales. 

\begin{figure}[t]
\hspace{14 mm}\includegraphics[width=2.3 in]{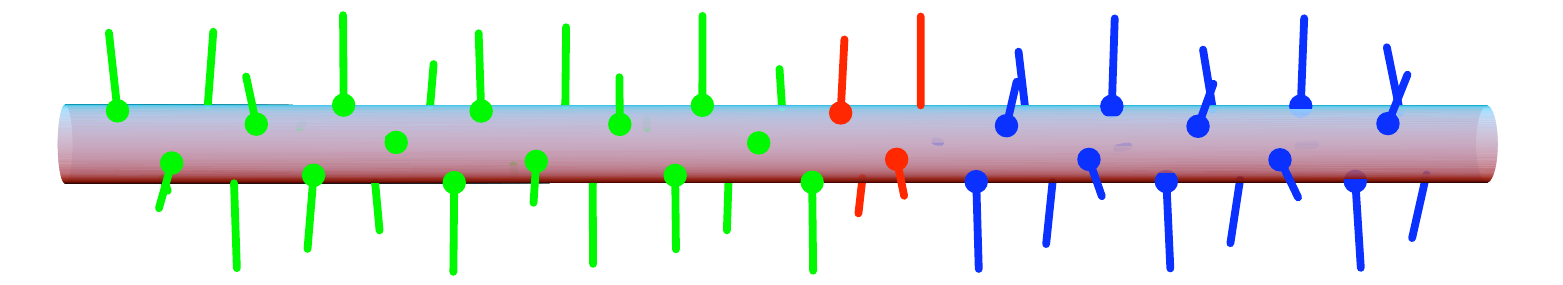}\vspace{5 mm}

\hspace{15 mm}\includegraphics[width=2.2 in]{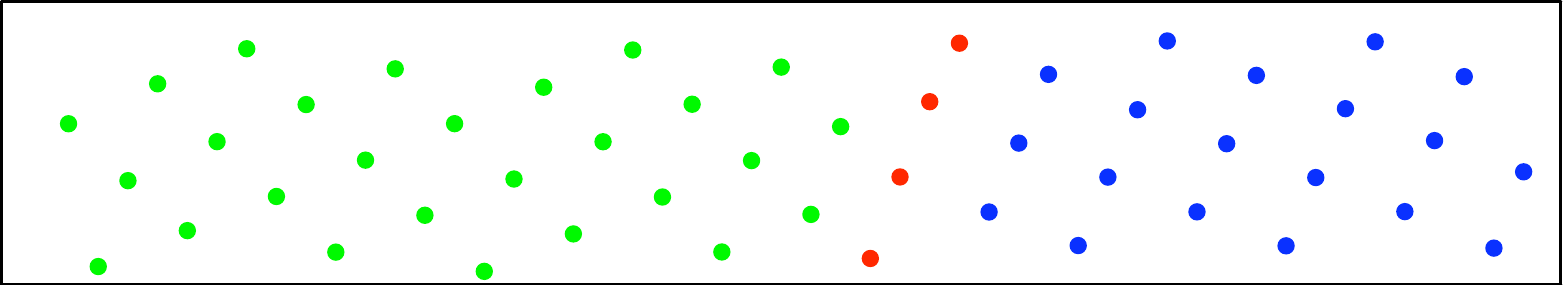}\vspace{5 mm}
\includegraphics[width=2.9 in]{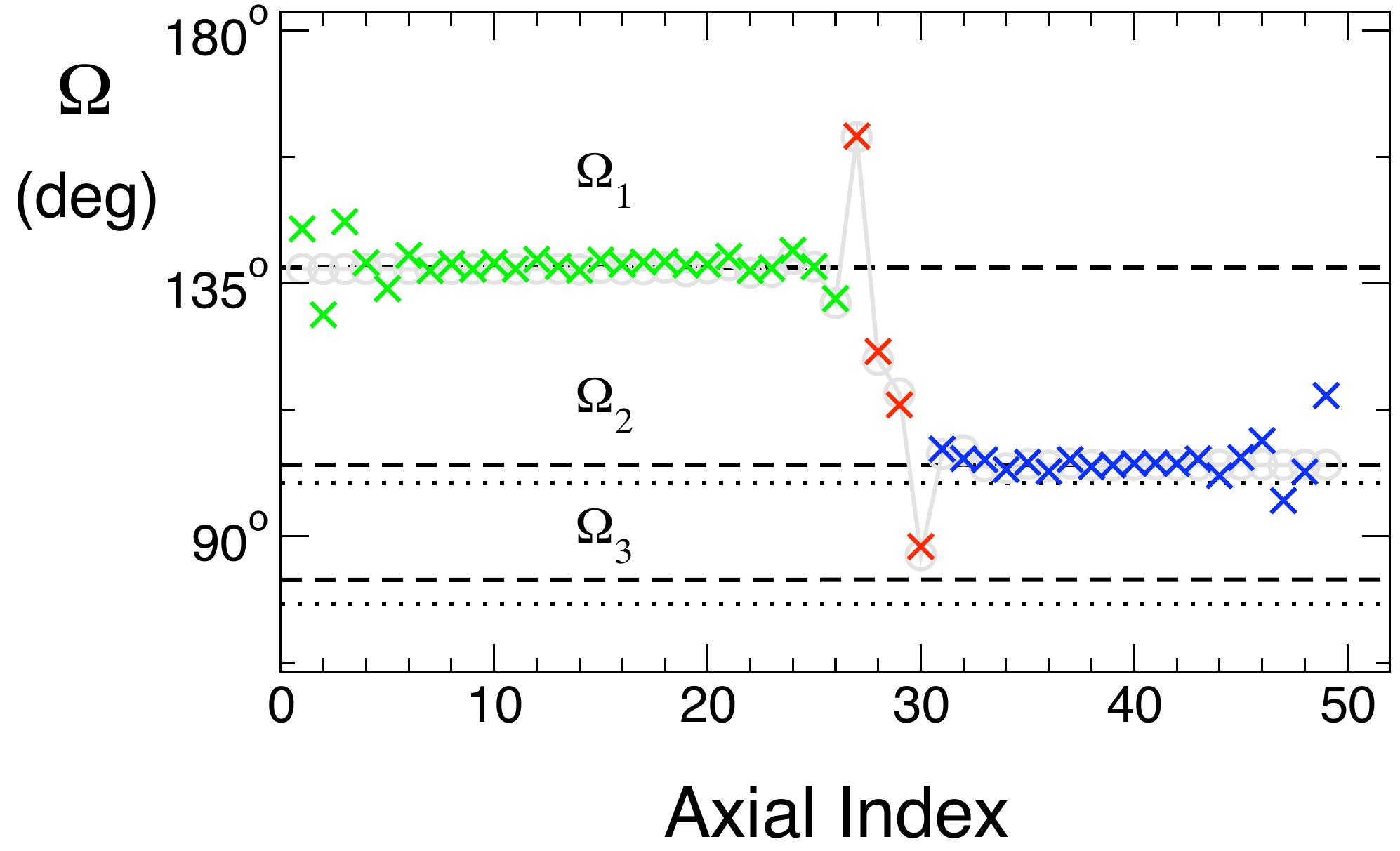}\vspace{5 mm}
\caption{ A kink between domains of  first and second phyllotaxis. From top to bottom: a 3-D rendering,  and its unwound 2-D Bravais lattice from numerical simulations. Below,  the same kink plotted as angle increments between successive magnets for the experimental system (crosses) and numerical simulation  (circles).}
\label{kink}
\end{figure}

\subsection{Results}

Figure~\ref{Fig2} reports the experimental results for both arrangements by plotting the measured angle between consecutive magnets. The more narrow (short-magnet) cactus self-organizes into the spirals  with divergence angles precisely reproducing those of first  phyllotaxis, $\Omega=\Omega_1$,  and second  phyllotaxis, $\Omega=\Omega_2$, as in Eq.~(\ref{Omega}). When the results are represented  in a 2-D lattice, as in the top of Fig.~\ref{Fig2}, parastichies can be  drawn. As parastichial numbers for $\Omega=\Omega_1$ we find the Fibonacci numbers $(2,3)$, and for  $\Omega=\Omega_2$, the Lucas numbers  $(3,4)$, as seen also in botany. 
The larger-radius system also forms first and second Phyllotaxis helices, as well as limited domains of third phyllotaxis [with $\Omega=\Omega_3$ and Lucas numbers $(1,4)$], bracketed by domains of second phyllotaxis. The
insets of Fig.~\ref{Fig2} show the magnetic interaction energy $V\left(\Omega\right)$ of the lattice obtained by interpolating measured values for the pair-wise magnet-magnet interaction, plotted as a function of divergence angle $\Omega$. As we cans see, local minima correspond to phyllotactic angles. 

\begin{figure}[t!!]
\begin{center}
\hspace{13 mm}\includegraphics[width=2.5 in]{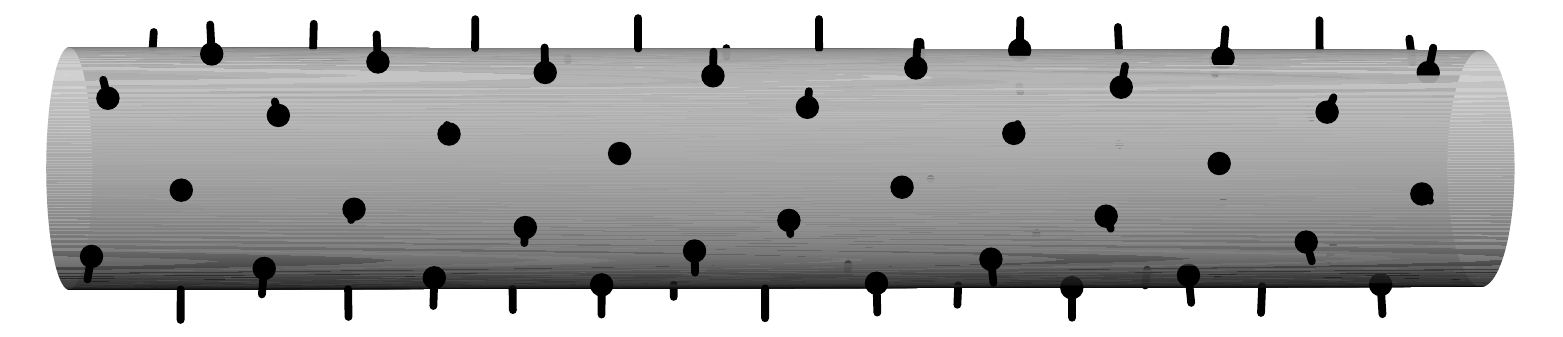}\vspace{4mm}

\includegraphics[width=3.2 in]{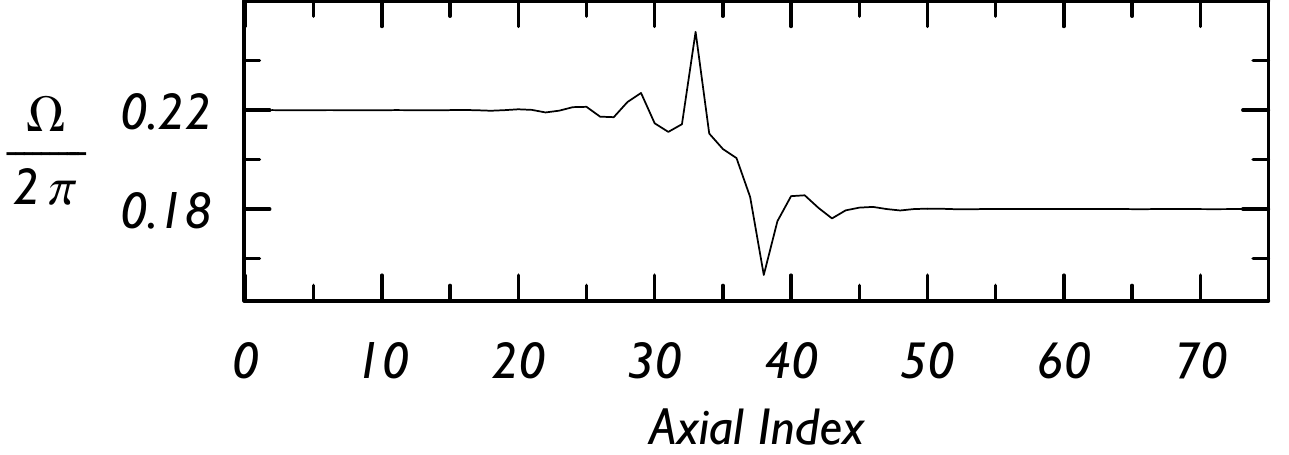}

\hspace{12 mm}\includegraphics[width=3.2 in]{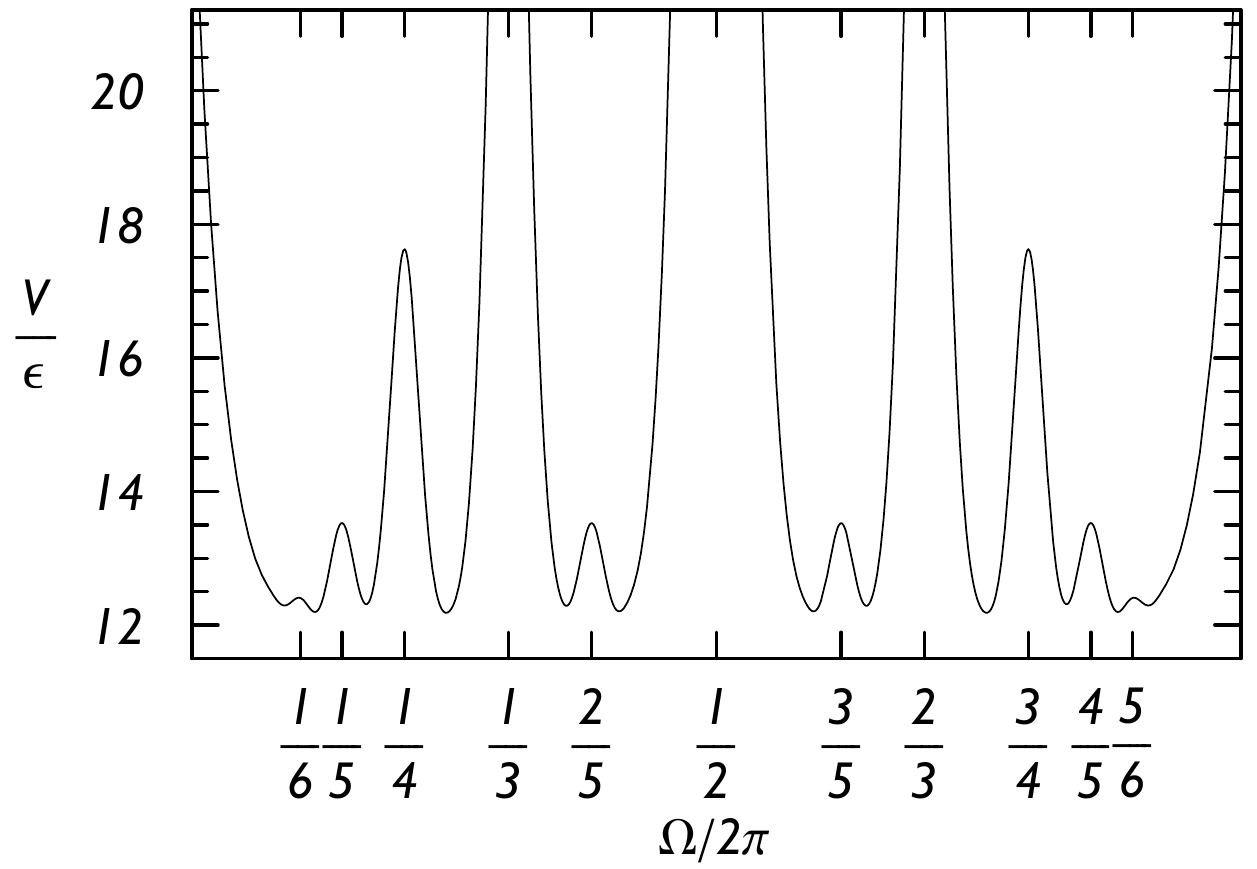}
\end{center}
\caption{A numerically calculated kink in a system of dipoles of high degeneracy (seven minima, $a/R=0.15$). The kink separates domains with parastichy numbers (4,5) and (5,6) and divergence angles of 1.38 and 1.13 radians. The top and middle panels give its three-dimensional rendering and angular shift $\Omega$ versus the axial magnet index. The two domains correspond to minima bracketed by peaks at $\Omega/2\pi = 1/4, 1/5$ and $\Omega/2\pi = 1/5, 1/6$ of the lattice energy, given in the bottom panel, where $\epsilon=p^2/a^2$, $p$ being the magnetic dipole.}
\label{kink2}
\vspace{-2mm}
\end{figure} 

\begin{figure}[t!]
\center
\includegraphics[width=3. in]{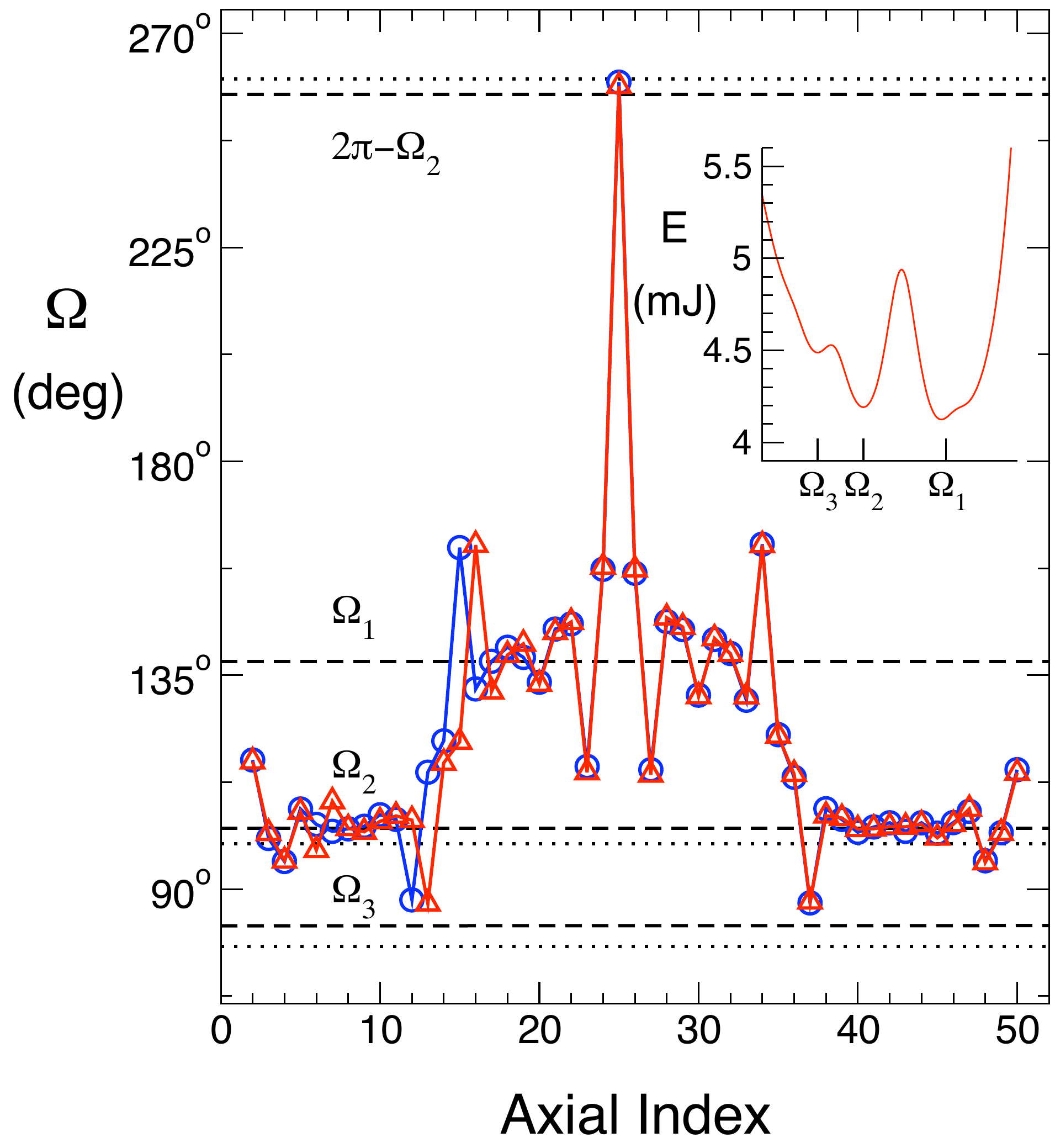}
\caption{Measured angular offsets between successive magnets for magnetic cacti with long  magnets, plotted versus magnet index, showing symmetric kink/anti-kink domain boundaries.  Dotted lines give the phyllotactic angles $\Omega_1$, $\Omega_2$, $\Omega_3$ and $2\pi-\Omega_2$ defined in the text. Dashed lines are minima of the magnetic lattice energy (inset) calculated by interpolating the measured pair-wise magnet-magnet interaction.}
\label{phantom}
\end{figure}
Figure~\ref{Fig2} also shows that in many instances the system fragments into two or three distinct domains whose domain walls always share a common parastichy, as seen in botany~\cite{kink,Adler}, and as expected in physics for a quasi-one-dimensional degenerate system. We have computed numerically  one such  transition via dynamical simulations in a velocity-Verlet algorithm, in the following way: we start from a crude static step-like kink as an initial condition, and allow it to radiate energy in the form of phonons waves until it stabilizes in a kink with superimposed vibrations; we then average this configuration over time,  to remove these residual oscillations. When the result is used as new initial conditions, it proves to be a static kink. Fig.~\ref{kink} reports our numerical results for a kink in a system whose size and interaction reproduces the physical realization of the magnetic cactus, along with the experimental data for such a kink. The match is essentially perfect, indicating that the dissipative (i.e. frictional) forces neglected in our model do not significantly affect the static configurations. We apply the same numerical procedure to calculate a kink in a system of larger degeneracy, among domains which are absent in our physical realization. We use a smaller $a/R$ ratio and a different interaction between particles (ideal dipole instead of physical dipole). The result shown  in Fig.~\ref{kink2} reproduces the same qualitative shape of the previous, lower degeneracy case. Similar kinks are present also in a fully unconstrained cactus, one in which the particles can move along the axis, and are found in early generations of our structural genetic algoritms (see below). 
Finally, these kinks can travel as novel topological solitons, with a rich phenomenology that is explained elsewhere~\cite{Nisoli_PRL, Nisoli_PRE}.

Finally, in the system with longer magnets, we occasionally found intriguing yet hard-to-interpret configurations that contain two nearly reflection symmetric domain boundaries. Fig.~\ref{phantom} reports two such configurations, measured in independent experimental runs. Although we do not have a firm explanation for these structures, we speculate that they form as frozen-in soliton waves that initiated symmetrically at both ends of the structure, upon release of the wound-up elastic energy during initial preparation. Indeed an analytical, continuum theory for phyllotactic solitons  which we have developed recently, and which explains results of the dynamical symulations   also  supports the existence of  similar frozen-in pulses~\cite{Nisoli_PRE}.

\section{Fully unconstrained cactus: structural genetic algorithm}

Our experimental apparatus is not fully unconstrained: the axial coordinates of the dipoles are fixed, and so only the azimuthal movement is allowed. While this is an huge improvement toward the original helical constraint of Levitov, many (most) physical systems that could manifest phyllotactic patterns do not posses such a lesser azimuthal constraint. To corroborate and extend our experimental results to a completely unconstrained system, we seek the energy minimum in a set of repulsive particles that can move {\it axially} as well as {\it angularly} on a cylindrical surface,  via a non-local numerical optimization. To this purpose, we developed a structural genetic algorithm.

\subsection{Genetic Algorithm}

A genetic algorithm is a method of optimization that mimics evolution to find the absolute minimum in a function which shows a large number of metastable minima. The coordinates of the energy functions are called genes, and a set of genes is a particular specification of value for those variables.   The routine typically starts with a set of ``parents'', or specific points in the domain of the energy function. At each stage of the routine, parents  ``mate''  to produce children via exchange of genes:  a subset  of the coordinates of each of the two configurations are swapped, therefore generating new points in the energy domain, called children. Each of those children is then locally relaxed to a minimum via a local search. The new population of parents and children undergoes genetic selection and only the fittest  (the lowest energy ones) form a new population. 

There are many different implementations of this general idea: care is taken not to lose genetic diversity during selection, to avoid a population of almost identical replicas; that is usually achieved with a more or less skilled genetic selection, which might retain less geneticaly fit individuals, and often by introducing mutations in the form of random alteration of the gene sequence, which would opefully prevent the routine from gettting stuck around a metastable region. Choice of parameterization of the structures (genes) and mating (crossover) is crucial to the performance of the algorithm. 

About fifteen years ago, Deaven and Ho~\cite{Deaven} introduced a so-called {\it structural} genetic algorithm, which proved particularly efficient in minimazing the energy of physical structures, as it allows for physical intuition in defining the genes and mating procedure. With it, they found the C$_{60}$ fullerene structure as a ground state of 60 carbon atoms interacting with suitable atomic potentials~\cite{Deaven} and solved the celebrated  Thomson problem of repulsive charges on a sphere~\cite{Morris}, a task quite similar to ours.   

\begin{figure}[t!!!]
\begin{center}
\hspace{14 mm}\includegraphics[width=3.1 in]{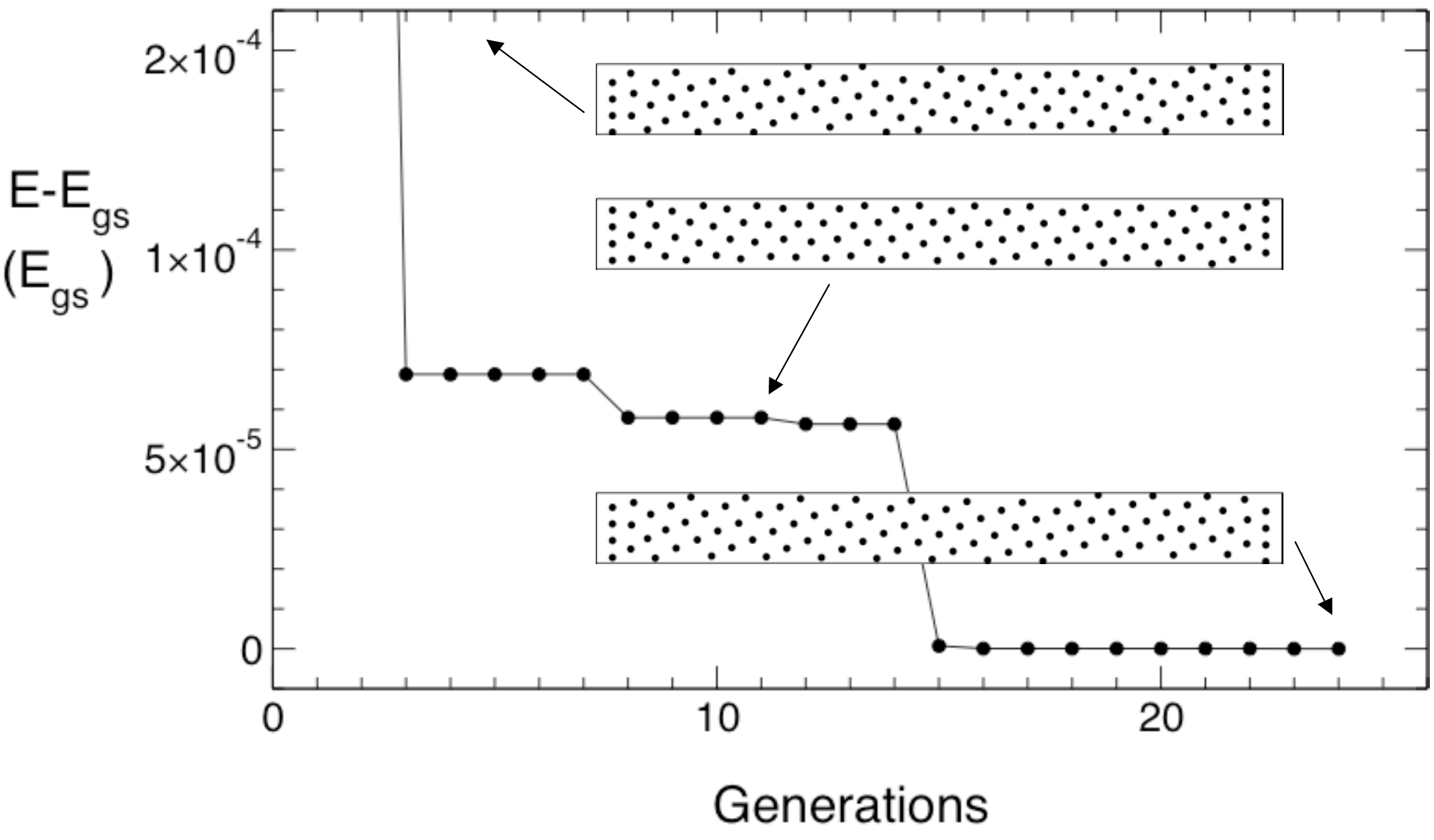}
\caption{Relative energy of the fittest member of the population in every generation. Early generations return very high energy configurations corresponding to disordered metastable states. Interrmediate generations show populations of  phyllotactic  domains separated by kinks between. Finally, the  algorithm converges to a single crystalline  domain in the bulk (deformations at the boundaries simply accomodate the system to the confining potential).}
\label{GAenergy}
\end{center}
\end{figure}

\subsection{Our Algorithm}

In our implementation we use a population of 10 members. Each member reppresents a configuration of 101 particles on the cylinder: more esplicitly, the genetic structure of each member $P_k$, $k=1,\dots,10$, of the population is a set of variables, or $P_k=\{ \theta_i,z_i \}^{101}_{i=1}$ which specifies, in cylindrical coordinates, the positions of the particles composing its structure. The particles interact via a pair-wise inverse quadratic repulsion $V=V_o (r_o/r)^2$, where $r$ {\it is the three dimensional distance between particles}; we introduce a confining potential in the form of an external axial square-well of width $L$, which sets the length of the cylinder, and hence the density. The choice of the pairwise interaction is not fundamental, as long as it is long ranged, repulsive, and well behaved~\cite{Nisoli_PRL}; our particular choice simply  speeds up the computation. 

We generate the first population  randomly. At each step,  we randomly couple mates, and excahge their genes employing the following mating procedure: we order the genes by increasing axial coordinates $z_1<z_2<\dots<z_{101}$ and swap the first $1<n<101$ genes, where $n$ is a random number, between randomly selected parents. The children obtained in this way are then relaxed to a stable structure via a standard conjugate gradient  algorithm. We then prepare the new generation by selecting the lowest energy individuals in the population of parents and children, yet making sure that the energy difference between members does not fall below a certain threshold, to preserve genetic diversity: when  new children cannot produce a new population of 10 in accordance with the energy threshold, we introduce mutations by randomly altering a certain number of members.  

\begin{figure}[t!!]
\begin{center}
\hspace{14 mm}\includegraphics[width=2.27 in]{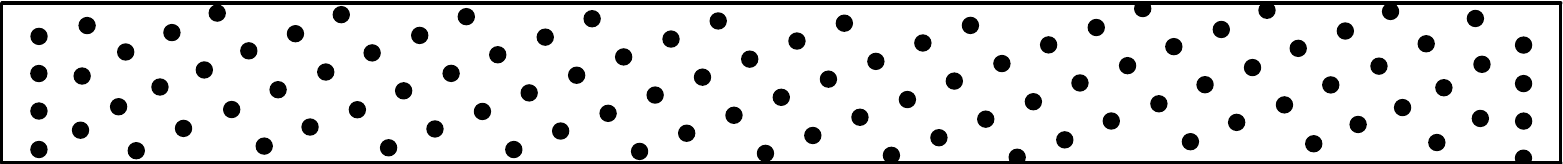}\vspace{5mm} 
\includegraphics[width=2.9 in]{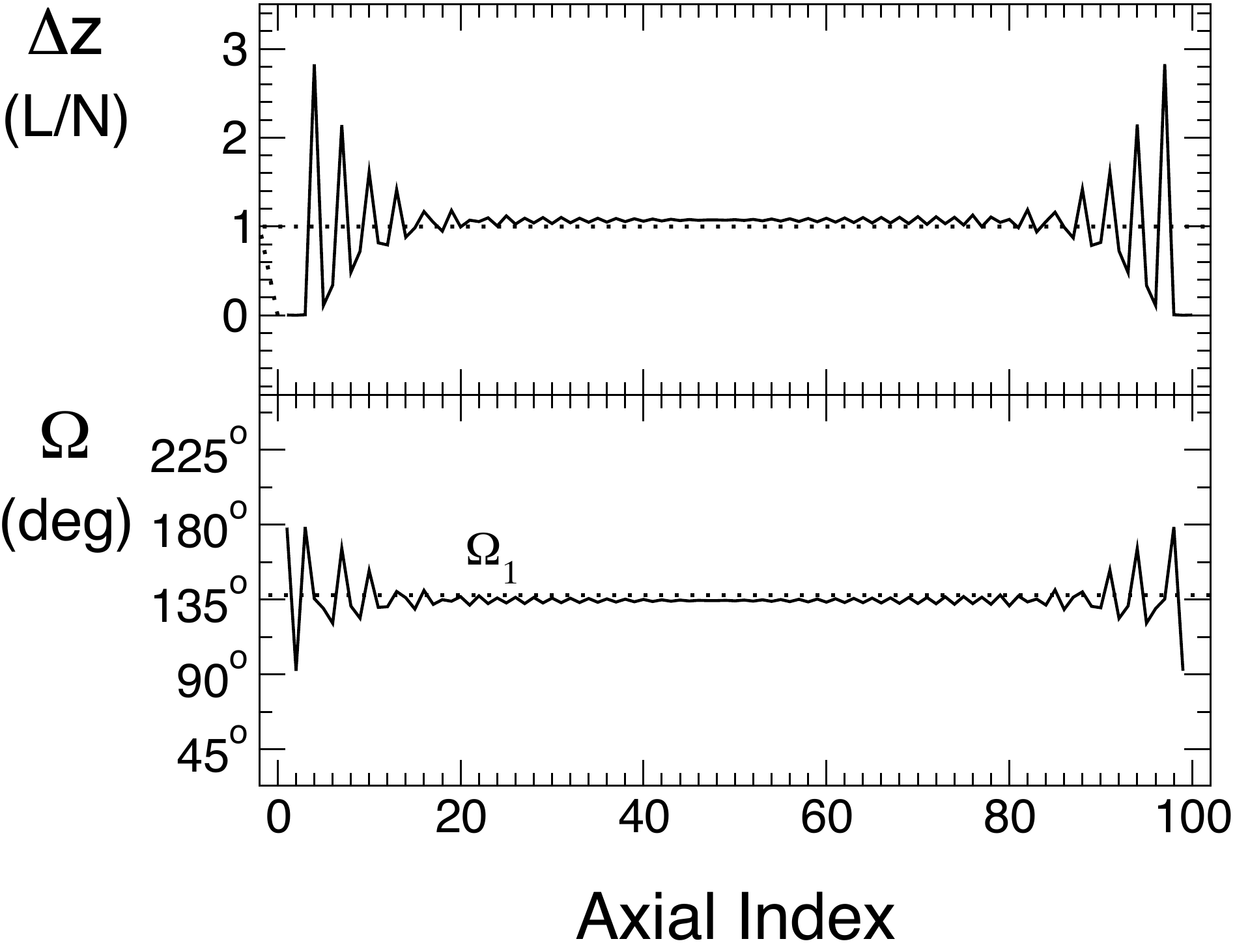}
\caption{Numerical optimization via structural genetic algorithm for $N=101$ repulsive particles [$V=V_o (r_o/r)^2$] constrained to a cylindrical surface of length $L$ and radius  $R=1.65 L/N$. The resulting 2-D Bravais lattice  has a nearly constant axial separation $\Delta z=z_{i+1}-z_i$ (top) and angular divergence $\Omega$  between successive particles (bottom), neglecting fringe effects at the border of the potential well. In the bulk, particles self-organize on a single spiral of divergence $\Omega=\Omega_1$. Oscillations at the boundaries are due to the effect of the confining potential.}
\label{Fig4}
\end{center}
\end{figure}

\subsection{Numerical Results}

During the structural evolution, the earliest populations contains metastable disordered states. Members of intermediate populations show kinks between domains of different divergence angle, configurations which are also seen experimentally. After fifteen to twenty generations, the algorithm typically converges to a single crystalline domain. 

Figure~\ref{GAenergy} reports the energy of the fittest member of the population at each generation in a typical run, showing a  punctuated-equilibrium evolution where the most-fit structure progressively decreases in energy in intermittent steps separated by plateaux. The final converged results form well-defined two-dimensional cylindrical crystals away from boundaries. 

Figure~\ref{Fig4} shows the crystalline structure to which the  algorithm converges, for $R=1.65 L/N$: a single spiral with $\Omega = \Omega_1$, as defined in Eqn.~(\ref{Omega}), corresponding to first phyllotaxis with parastichies $(1,2)$.  A plot of $\Delta z=z_{i+1}-z_i$ returns the value $L/N$ in the bulk, which implies a single generative spiral. This choice of $R N/L$ corresponds to a density close to that of our experimental apparatus. 

\section{Multijugate phyllotaxis}

\begin{figure}[t]
\begin{center}
\hspace{7 mm}\includegraphics[width=2.8 in]{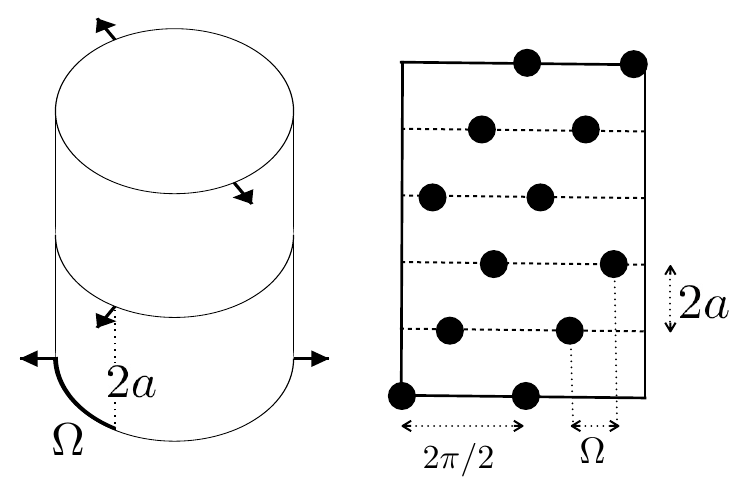}\vspace{6 mm}

\hspace{-4.55 mm}\includegraphics[width=3.15 in]{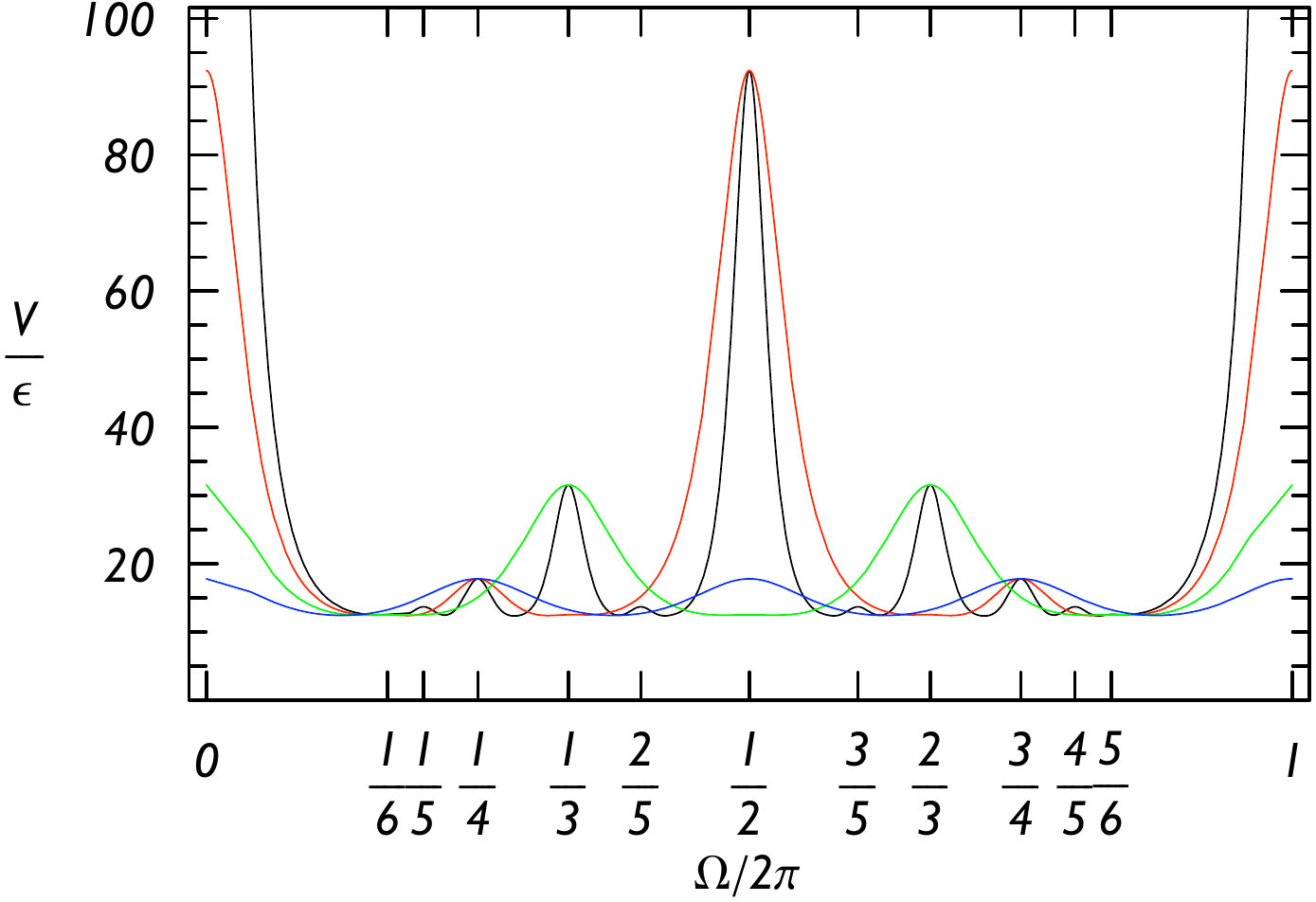}
\caption{ Top: 3-D schematics and 2-D lattice for a 2-jugate configuration. Bottom: Lattice energy versus angular offset for n-jugate configurations in a system of repulsive dipoles  ($\epsilon=p^2/a^2$, where $p$ is the magnetic dipole) for 1-jugate (black),  2-jugate (red), 3-jugate (green), and 4-jugate (blue).}
\label{multi}
\end{center}
\end{figure}

For highly degenerate systems the genetic algorithm returns configurations with more than one generative spiral, corresponding  to what in botany is called multijugate phyllotaxis~\cite{Jean}.  We have seen before that helices make cylindrically symmetric lattices. On the other hand, every cylindrically symmetric lattice can be decomposed into a suitable number of equispaced generative spirals~\cite{Jean, Bravais}. That is accomplished by discretizing  the  cylinder along its axis into equally spaced rings and then assigning at each ring $n$ sites, equally spaced  and separated by a $2\pi/n$ angular shift. As before, each ring is shifted consecutively by a divergence angle $\Omega$. The case $n=2$ is shown at the top of Fig.~\ref{multi}. The case $n=1$ is shown in our experimental arrangement of Fig.~\ref{Fig1}. 

By decomposing the n-jugate cylindrical lattice into $n$ lateral replicas of  single-spiral lattice, as in Fig.~\ref{Lattice}, the reader is easily convinced  that multijugate phyllotaxis reduces to the previously described monojugate case. All the  considerations above apply, provided that one now takes  the periodicity to be $2\pi/n$, and the distance between rings to be $n a$ (with, as before, $a=L/N$). It follows that  a n-jugate configuration will have local maxima in $2\pi i/n$, $i=1\dots n$ and, following the discussion of section II, one finds that there will be other local maxima corresponding to angles $2\pi/n\times i/j$ when $j \le [[J/n]]$, and $J$ is the maximum rank given by Eq.~\ref{J}.

Note now that if two multijugate lattices of jugation $n$, $n'$ have a peak in the commensurate angle $2\pi i/j$, then the energy of the peak is the same, as is shown in Fig.~\ref{multi}, bottom,  which  compares the plots of the energy of such an arrangement for different values of $n$. In fact both configurations correspond to particles facing each other after  $j/n$ and $j/n'$ rings, and therefore at the same distance $na\times j/n= n'a\times j/n' =j a$, independent of $n$ or $n'$. For small $n$  the minima in the energy  of n-jugate configurations essentially degenerate with the monojugate one previously explored. For large $n$ they have higher energy. If $a/R$ is small enough, the threshold is  $n>J$, as interaction between particles on the same ring become comparable to those facing in the minimal monojugate peak.

\section{Conclusion}

We have studied the lowest energy configurations of repulsive particles on cylindrical surfaces, both experimentally and numerically. We have found that they correspond to the spiraling lattices seen in the  phyllotaxis of living beings, both monojugate and multijugate. By establishing experimentally and numerically that phyllotactic point lattices are ground states in the very general geometric scenario of unconstrained repulsive particles on cylinders, we have opened the study of phyllotaxis to a much wider range of annealable physical systems where the particles could be electrons, adatoms, ions, dipolar molecules, nanoparticles, etc. constrained by external potentials. 

Unlike plants, these multifarious, non-biological Phyllotactic systems could access various degrees of dynamics, providing new phenomenology well beyond that available to over-damped, adiabatic botany. We have reported elsewhere~\cite{Nisoli_PRL} on the dynamical richness of this physical phyllotaxis, including classical rotons and a large family of novel, inter-converting topological solitons.

\end{document}